\documentclass[a4paper,11pt]{article}
\usepackage[utf8]{inputenc}
\usepackage{mathpazo}
\usepackage[margin=2.5cm]{geometry}
\usepackage{hyperref}
\usepackage[frozencache=true,cachedir=.]{minted}
\usepackage{amsmath}

\newtheorem{theorem}{Theorem}[section]
\newtheorem{definition}{Definition}[section]
\newtheorem{note}{Note}[section]

\usepackage[style=authoryear]{biblatex}
\title{UFL Dual Spaces, a proposal}
\author{David Ham}
\date{January 2021}

\begin{document}

\maketitle

\begin{abstract}
    This white paper highlights current limitations in the algebraic closure Unified Form Language (UFL). UFL currently represents forms over finite element spaces, however finite element problems naturally result in objects in the dual to a finite element space, and operators mapping between primal and dual finite element spaces. This document sketches the relevant mathematical areas and proposes changes to the UFL language to support dual spaces as first class types in UFL.
\end{abstract}

\section{Introduction}

UFL is the Unified Form Language. It is a language used to describe multilinear forms. Typically these forms are to be used in a finite-dimensional variational problem, specifically a finite element problem. UFL has not, hitherto, concerned itself with the spaces to which forms belong or into which they map: this has been the concern of the assembler, such as Firedrake or FEniCS and, to some extent, the linear algebra backend (PETSc in the relevant cases, though other choices would be possible). These spaces involve the duals to the spaces in which UFL objects such as Coefficients and Arguments are defined. The absence of dual spaces in UFL means that the language is not closed: there are operations visible to the UFL user which result in objects that are not UFL objects, or which rely on bespoke functions where a more closed system would have a first class operator. Examples of this include:

\begin{enumerate}
    \item Interpolation, which is a form whose range space is the dual of the space being interpolated into.
    \item Nonlinear constitutive laws for materials such as ice. In these cases, UFL-based systems currently rely on an unnecessary Galerkin projection to represent an operation which is not a differential operator.
    \item The adjoint, which currently has to rely on intricate reasoning to work out which objects are primal or dual, where this should be obvious from their type.
\end{enumerate}

This document is an outline proposal for the inclusion of dual spaces and related objects in UFL. It has particular implications for ongoing work in Firedrake on interpolation and operators external to the UFL system.

\section{Preliminaries}

We will assume that there is a domain $\Omega \subset \mathbb{R}^n$ for some $n\ge0$ which is the region on which we will calculate.

\begin{definition}[Function space]
    A \emph{function space} is a vector space over a field K comprising functions $\Omega\rightarrow K^m$.
\end{definition}
That is to say, if $V$ is a function space and $f\in V$ then $f$ associates a value in $K^m$ with each point in the domain $\Omega$. The field $K$ will be either $\mathbb{R}$ or $\mathbb{C}$. 

\begin{definition}[linear $k$-form]\label{kform}
    A \emph{linear $k$-form} is a linear mapping from $k\ge0$ vector spaces to $K$. That is, a linear $k$-form is a function:
    \begin{equation} 
    V_{k-1} \times V_{k-2} \times \ldots \times V_{0} \rightarrow K
    \end{equation} 
    for suitable vector spaces $V_0\ldots V_{k-1}$. 
\end{definition}

\begin{note}
    The argument numbering in definition \ref{kform} is deliberately from right to left. This reflects the numbering of arguments in UFL. It also caters for a difference in notation between forms and linear algebra: the assembled tensors are the transpose of the form, so the function spaces number from left to right when we consider the eventual linear algebra operations.
\end{note}

\begin{definition}[sesquilinear $k$-form]
    A \emph{sesquilinear $k$-form} is a mapping from $k\ge0$ vector spaces to $K$ which is conjugate linear (also termed antilinear) in the final argument (for $k>0$), and linear in the others. 
\end{definition}

\begin{definition}[Inner product]
    An inner product defined on a vector space $V$ is a sesquilinear 2-form $V\times V\rightarrow K$. Which is Hermitian symmetric and positive definite. That is, for any $u, v \in V$
    \begin{enumerate}
        \item $\langle u, v\rangle = \overline{\langle v, u \rangle}$
        \item $\langle u, u\rangle \geq 0$
    \end{enumerate}
\end{definition}

\begin{definition}[Hilbert space]
    A Hilbert space is a complete normed vector space equipped with an inner product. All of the function spaces we are interested in are Hilbert spaces.
\end{definition}

\begin{definition}[$L^2$ inner product]
    The $L^2$ inner product between two functions $u, v: \Omega\rightarrow K$ is given by:
    \begin{equation} 
        \langle u, v \rangle_{L^2} = \int_{\Omega} u\cdot \overline{v}\ \mathrm{d}x 
    \end{equation} 
\end{definition}

\begin{definition}[$L^2$]
    The function space $L^2(\Omega)^n : \Omega\rightarrow K^n$ is the Hilbert space of functions where the norm induced by the $L^2$ inner product is finite:
    \begin{equation} 
        u\in L^2(\Omega)^n \Leftrightarrow \sqrt{\langle u,u\rangle_{L^2}}\in \mathbb{R}
    \end{equation} 
\end{definition}

\begin{definition}[Dual space]\label{dualspace}
    If $V$ is a vector space then $V^*$, the dual space to $V$, is the space of bounded linear functionals $V\rightarrow K$.
\end{definition}
For the finite-dimensional spaces with which we are concerned, it is easy to establish the following:
\begin{theorem}
If $V$ is a finite-dimensional vector space, then $|V| = |V^*|$
\end{theorem}

\begin{theorem}[Dual basis]
If $\{\phi_i\}$ is a basis for a finite vector space $V$ then there exists $\{\phi_i^*\}$ a basis for $V^*$ such that:
\begin{equation} 
    \phi_i^*(\phi_j) = \delta_{ij}
\end{equation} 
The basis $\{\phi_i^*\}$ is termed the \emph{dual basis}.
\end{theorem}
Where it is necessary to make the distinction, we will refer to the space to which a dual space is dual as the \emph{primal space} and its basis as the \emph{primal basis}. 
\begin{definition}{Reflexive vector space}
    A reflexive vector space is one for which $(V^{*})^{*}$ is isomorphic to $V$ under the canonical map. That is, we can identify $(V^*)^*$ and $V$.
\end{definition}
\begin{theorem}
    All finite-dimensional vector spaces are reflexive.
\end{theorem}
This is a completely intuitive result. If $u\in V$ and $v\in V^*$ then this induces a $u'$ in $(V^*)^*$ given by:
\begin{equation}
    u'(v) = v(u)    
\end{equation}
We simply identify $u'$ and $u$. Since the dimensions of the spaces match and the functionals are linear, this identification is an isometric isomorphism. In other words we can't distinguish the spaces so the identification is valid.

\begin{theorem}
    All Hilbert spaces are reflexive.
\end{theorem}

\section{Functional and tensor product notation}

In definition \ref{kform} and elsewhere, we use a notation for multilinear forms which takes a Cartesian product of spaces and returns a scalar. For example for a 2-form:
\begin{equation}
    V_1 \times V_0 \rightarrow K
\end{equation} 
This notation emphasises that the form is a function which takes two vectors and returns a scalar. However we could just as well Curry this form and write:
\begin{equation}  
    V_1 \rightarrow V_0 \rightarrow K
\end{equation} 
This notation is left-associative, so this means:
\begin{equation} 
    V_1 \rightarrow (V_0 \rightarrow K)
\end{equation} 
In other words, a 2-form is a function which takes a vector and returns a 1-form.

\section{Dual spaces and $k$-forms}

In this section we restrict our attention to real-valued function spaces so as not to have to consider where all the conjugates should be placed. We will come back to the complex valued case.

It follows immediately from definitions \ref{kform} and \ref{dualspace} that the space of linear 1-forms on vector space $V$ is the same thing as $V^*$. This is nothing else than writing:
\begin{equation} 
    V^* = V\rightarrow \mathbb{R}
\end{equation} 
A direct consequence of this is the following equivalence:
\begin{equation} 
    V_1 \times V_0 \rightarrow \mathbb{R} = V_1 \rightarrow V_0^*
\end{equation}
This says that every form can equivalently be thought of as an operator mapping into the dual of its 0-number space.

\section{The interpolation operator}

Usually we are concerned with vector spaces which are finite-dimensional subspaces of larger, typically infinite dimensional, spaces. In the finite element method, we often define the basis of these finite spaces by defining the space and choosing the dual basis. The primal basis can then be derived by solving a linear system (see, for example, \href{https://finite-element.github.io/2_finite_elements.html}{the finite element course}). Let's briefly examine the dual basis functions, also known as nodes. These are functionals that take in a function and (linearly) produce a scalar. What do such objects look like? The simplest case is probably the point evaluation functional, or Dirac delta distribution. For a function $v: \Omega\rightarrow\mathbb{R}$, the point evaluation functional at the point $x$ is given by:
\begin{equation}
    \delta_x(v) = v(x)
\end{equation}
Other common functionals include the integral of a function over one mesh cell, or the component of a vector value in a particular direction at a particular point. All of these have in common that they are trivially extensible to a much larger space of functions than the finite element function space whose dual they define. For example, point evaluation functionals can be evaluated for any space of functions whose members are well-defined at the points in question. This is useful because int is a very common requirement to approximate some function from a more general space in a finite element function. This most commonly occurs when a forcing function or reference solution is known either from external (physical) data, or from an analytic expression. Suppose $v\in V$ and that $\{\phi_i\}$ is a basis for $V$, with $\{\phi_i^*\}$ the corresponding dual basis. The for a function $f:\Omega\rightarrow\mathbb{R}$ such that the dual functionals are defined, the interpolation of $f$ into $v$ is given by:
\begin{equation}
    \hat{f} = \hat{f}_i \phi_i 
\end{equation}
where:
\begin{equation}
    \hat{f}_i = \phi_i^*(f)
\end{equation}
In other words, the $i$-th basis coefficient is the evaluation of the $i$-th dual basis functional on the function to be interpolated. We can call this operator $I_V(f)$.
\begin{note}
    If $f\in V$ then it is straightforward to show that $I_V(f)=f$. That is to say, interpolation is a projection.
\end{note}
In fact, the interpolation operator is nothing other than the evaluation of the dual basis. We could instead define an operator, a generalisation of the delta distribution, which takes a functional $v\in V^*$ and applies it to a function $u\in U$ where $U$ is a function space on which the functionals in $V^*$ are defined. That is to say:
\begin{equation}
    \delta(u, v) = v(u)
\end{equation}
Notice that $\delta: U\times V^*\rightarrow \mathbb{R}$. Since $V^*$ is itself a vector space, this makes $\delta$ a 2-form, albeit a slightly unusual one. We would be more used to forms in $U\times V\rightarrow\mathbb{R}$.

\section{UFL Forms}

UFL is the Unified Form Language \parencite{alnaes2014} also described in \cite{homolya2018}. As one might expect, it provides support for $k$-linear forms. UFL is a symbolic algebra language. In common with other such languages, it provides for symbols to stand for unknown values, including the arguments to $k$-linear forms. UFL is a purely symbolic language which problem solving environments such as Firedrake and FEniCS extend by subclassing in order to attach numerical data and actually solve problems. In understanding how UFL objects do (or should) behave, it is sometimes necessary to consider how they would be integrated into the problem solving environment, and especially what should happen when a form is assembled.

In an attempt to avoid replicating the documentation of UFL, we will start here from UFL \mintinline{python}|FunctionSpace| and gloss over the cells, meshes, and finite elements from which these are constructed.

If \mintinline{python}|V| is a \mintinline{python}|ufl.FunctionSpace| then \mintinline{python}|c = ufl.Coefficient(V)| defines a \emph{known} function in \mintinline{python}|V|. Conversely, \mintinline{python}|a = ufl.Argument(V, 0)|\footnote{Users more frequently write \mintinline{python}|a = ufl.TestFunction(V)|, but that is simply syntactic sugar. A test function is just argument 0 to a form.} defines a placeholder symbol for an \emph{unknown} function in \mintinline{python}|V|.
Consequently, 
\mint{python}|f_0 = c * dx| 
is a symbolic expression for the integral of \mintinline{Python}|c| over the domain and is a real value. UFL calls this a 0-form. In contrast, 
\mint{python}|f_1 = a * dx|
represents the integration of the unknown function \mintinline{python}|a| over the domain. It's therefore a 1-form, or a function in $V\rightarrow\mathbb{R}$.

In order to make the argument order clear, we introduce a second function space, \mintinline{python}|W|. We  define \mintinline{python}|b = ufl.Argument(W, 1)|\footnote{Equivalently, \mintinline{python}|b = ufl.TrialFunction(W)|.} and create the simplest 2-form:
\mint{python}|f_2 = c * b * dx|
This is a function in $W\times V\rightarrow\mathbb{R}$. Notice that, by convention, argument 1 comes before argument 0.

\subsection{Assembly}

The problem solving environment takes symbolic UFL forms and evaluates them. In Firedrake and FEniCS, the operation which does this is called assembly. The problem solving environment encodes in its subclass of \mintinline{python}|ufl.FunctionSpace| a basis for the function space. We will write ${\phi_i}$ for the basis for $V$ and ${\psi_i}$ for the basis for $W$. The problem solving environment will also implement its own subclass of \mintinline{python}|ufl.Coefficient|, in Firedrake and FEniCS this is called a \mintinline{python}|Function|. If \mintinline{python}|c| is a \mintinline{python}|Coefficient|, then mathematically:
\begin{equation}
    c = c_i\phi_i
\end{equation}
The consequence of this is that the core of the implementation of a \mintinline{python}|Function| is a reference to a \mintinline{python}|FunctionSpace| (which encodes $\{\phi_i\}$) and a vector of values $c_i$. The latter give the \mintinline{python}|Function| its numerical value, while the former gives that value a mathematical meaning.

If we execute:
\begin{minted}{python}
    g = assemble(f_0)
\end{minted}
Then the integral is evaluated and $t_0$ is a scalar value. Mathematically, nothing has really happened: $g$ and $f_0$ represent the same scalar value. Computationally, however, we have gone from a symbolic object which evaluates to a scalar, to the scalar itself.

So, what happens in the following case?
\begin{minted}{python}
    h = assemble(f_1)
\end{minted}
\mintinline{python}|f_1| contains the argument, or unknown \mintinline{python}|Function|, \mintinline{python}|a|. In other words, we need to evaluate
\begin{equation}
    \begin{split}
        h(a) &= \int a\,\mathrm{d}x\\
        &= \int a_i \phi_i\,\mathrm{d}x
    \end{split}
\end{equation}
for unknown values $a_i$. However the $a_i$ are scalar so they can be lifted out of the integral:
\begin{equation}
    \int a\,\mathrm{d}x = \int \phi_i\,\mathrm{d}x\ a_i
\end{equation}
Observe that we can now evaluate all the integrals, as they no longer contain unknown values. 
Next, we use the fact that $\phi^*_i(\phi_j) = \delta_{ij} = \mathrm{I}$, the identity matrix. Hence:
\begin{equation}
    \begin{split}
        h(a) &= \int \phi_i\,\mathrm{d}x\ \mathrm{I}_{ij} a_j\\
        &= \int \phi_i\,\mathrm{d}x\ \phi^*_i(\phi_j) a_j\\
        &= \int \phi_i\,\mathrm{d}x\ \phi^*_i(a_j\phi_j)\\
        &= \int \phi_i\,\mathrm{d}x\ \phi^*_i(a)\\
    \end{split}
\end{equation}
We used the linearity of the dual basis functionals for the penultimate line. We have now expressed $h$ in terms of the basis for $V^*$: $h = h_i \phi^*_i$ where:
\begin{equation}
    h_i = \int \phi_i\ \mathrm{d}x
\end{equation}
The consequence of this is that the value of $h$ is encoded in a vector of numbers, of the same size as that which encodes $c$. Now one could say that the only purpose of assembly is to create a linear system to be solved, so the vector of values is the only thing which is required. This is the current FEniCS approach. This approach is problematic because it neglects the fact that assembled operators can and are used by users in other contexts. 

What is missing, of course, is a reference to the dual basis $\{\phi^*_i\}$. The challenge is that UFL does not have a class encoding $\{\phi^*_i\}$, and hence also does not have a suitable class for $h$. The Firedrake approach is to appeal to the natural isomorphism between $V$ and $V^*$ and simply make \mintinline{python}|h| a \mintinline{python}|Function|. This has the advantage of preserving the link between the numerical values and the function space, but at the expense of incorrectly labelling $t_1$ as being in $V$. What we should really do is associate the list of coefficients $h_i$ with an object encoding the dual basis, in a manner analogous to the association between a list of coefficients and the basis for a function space in a \mintinline{python}|Function|. 

To illustrate the manner in which the assembled form encodes the same information as the symbolic form, observe that:
\begin{equation}
    h(c) = f_1(c) = \int c\,\mathrm{d}x = f_0 = g
\end{equation}
However the evaluation of $h(c)$ is actually accomplished like this:
\begin{equation}
    \begin{split}
        h(c) &=  h_i \phi^*_i(c_j\phi_j)\\
        &= h_i \phi^*_i(\phi_j) c_j\\
        &= h_i \delta_{ij}c_j\\
        &= h_i c_i
    \end{split}
\end{equation}
So the assembly has turned the evaluation of an integral into the dot product between two vectors.

\subsection{UFL dual spaces and cofunctions}

The previous section suggests an outline of at least some missing UFL functionality. There should be an object representing the dual to a function space (\mintinline{python}|DualFunctionSpace| or maybe just \mintinline{python}|DualSpace|?) Presumably one would add a property to \mintinline{python}|ufl.FunctionSpace| to access the dual, and of course \mintinline{python}|DualSpace| would also have this property but it would point back to the primal space. Maybe \mintinline{python}|ufl.FunctionSpace| and \mintinline{python}|ufl.DualSpace| have a common parent class \mintinline{python}|ufl.VectorSpace|.

We also need to be able to define a concrete member of $V^*$. Consistency with the existing UFL usage would suggest that this class would be \mintinline{python}|ufl.Cocoefficient| but this is a ridiculous and cumbersome name so perhaps \mintinline{python}|ufl.Cofunction| would be better. Firedrake and FEniCS subclasses of \mintinline{python}|ufl.Cofunction| would, in a manner analogous to \mintinline{python}|Function| have both a vector of values and a reference to the \mintinline{python}|ufl.DualSpace|. The question of which operations should be valid for a cofunction requires more thought, though these would certainly include the basic vector operations of addition and multiplication by a scalar.

\subsection{Cofunctions are forms}

A cofunction is mathematically just a 1-form. The distinction between this and an existing UFL 1-form is that the latter is a symbolic expression that represents a 1-form, while the former is a numerical 1-form represented directly in the dual basis. Since they are mathematically objects of the same type, they should be arithmetically interchangeable. In particular, if \mintinline{python}|f = ufl.Cofunction(V.dual)|  and \mintinline{python}|v = ufl.Argument(V, 0)| then the following makes mathematical sense as a form and should be allowed:
\begin{minted}{python}
    g = f + v*dx
\end{minted}
This possibly implies that cofunctions intrinsically have an argument drawn from the relevant primal space. Assembly of \mintinline{python}|g| can be achieved by simply defining assembly of a cofunction as the identity operation.

\subsection{2-forms}

What do we get from:
\begin{minted}{python}
    A = assemble(f_2)
\end{minted}
Mathematically this is defined as follows:
\begin{equation}
    A_{ij} = \int \phi_i \psi_j\ \mathrm{d}x
\end{equation}
That is to say, $A$ is a matrix. Recall that $\{\phi_i\}$ is the basis for $V$, which is the function space of argument 0 in this case, and $\{\psi_j\}$ is the basis for $W$, which is argument 1. This means that the assembled operator has its arguments in ascending order, while the form notation has them in descending order. This is a frequent cause of confusion about which we will have to remain aware. FEniCS and Firedrake both agree that assembling a 2-form produces a matrix, though UFL does not have a corresponding class, so currently matrices are outside its scope. However, a matrix is an assembled 2-form in just the same way that a cofuncion is an assembled 1-form, so the logical conclusion would be the creation of a \mintinline{python}|ufl.Matrix| class which would behave as an assembled 2-form, and would have algebraic properties analogous to a cofunction. A matrix would presumably have two intrinsic arguments.

\section{The delta operator and dual space arguments}

Let's now return to the interpolation operator. If we wish to incorporate it into UFL as a first class citizen then we need to take our extensions of the language just a little further. Let's suppose we can define arguments in dual spaces. These are mathematical objects of a different kind to the arguments we have already met: they are unknown members of $V^*$ rather than $V$. We suppose, therefore, that they are of a new type: \mintinline{python}|ufl.Coargument|. In order to minimise syntactic clutter, one might imagine overloading the object constructor such that \mintinline{python}|ufl.Argument(V.dual, 0)| is valid and returns a \mintinline{python}|ufl.Coargument|.

This enables us to define a new form \mintinline{python}|ufl.Delta| such that if \mintinline{python}|e| is a UFL expression and \mintinline{python}|v_ = ufl.Argument(V.dual, 0)| then \mintinline{python}|delta(e, v_)| is a symbolic expression for the interpolation of \mintinline{python}|e| into $V$. If we assume that \mintinline{python}|e| does not contain any arguments and execute:
\begin{minted}{python}
    f = assemble(Delta(e, v_))
\end{minted}
Then \mintinline{python}|f| will be a \mintinline{python}|Function| in $V$ representing the result of this interpolation. Conversely, if \mintinline{python}|u = Argument (W, 1)| then:
\begin{minted}{python}
    A = assemble(Delta(u, v_))
\end{minted}
results in a matrix which is the interpolation operator from $W$ into $V$. One can also construct the adjoint of this operator by reversing the argument positions:
\begin{minted}{python}
    v = ufl.Argument(W, 0)
    u_ = ufl.Argument(V.dual, 1)
    AT = assemble(ufl.Delta(v, u_))
\end{minted}
As a particular case of interest, if we write:
\begin{minted}{python}
    u = ufl.Argument(V, 1)
    v_ = ufl.Argument(V.dual, 0)
    I = assemble(ufl.Delta(u, v_))
\end{minted}
Then \mintinline{python}|I| is the identity matrix for $V$

For completeness, we can also address the other cases. If we write:
\begin{minted}{python}
    f = firedrake.Cofunction(V.dual)
    c = firedrake.Function(W)
    t = ufl.Delta(c, f)
    s = assemble(t)
\end{minted}
then $t$ is a symbolic expression for $f(c)$ and $s$ is the scalar which results from evaluating it. Indeed, one might want to implement the \mintinline{python}|__call__| special method on \mintinline{python}|ufl.Cofunction| and \mintinline{python}|ufl.Coargument| so that the syntax \mintinline{python}|f(c)| works directly. Note that this would simply return \mintinline{python}|ufl.Delta(c, f)|. \mintinline{python}|ufl.Delta| is still needed as the object which encodes the symbolic operation of dual evaluation.

Finally, if we write:
\begin{minted}{python}
    f = firedrake.Cofunction(V.dual)
    u = ufl.Argument(W, 0)
    g = assemble(ufl.Delta(u, f))
\end{minted} 
Then $g\in W^*$ is the interpolation of $f$ into $W^*$. (Note that Cofunction interpolation is the transpose of function interpolation).

\section{Adjoint}

The  dual or adjoint of a 2-form is the conjugate of that form with its arguments relabelled in reversed order. This carries over to forms which map from or to dual spaces. If we take the example of the delta form again, then:
\begin{equation}
    \delta: V \times W^* \rightarrow \mathbb{R}.
\end{equation}
Equivalently:
\begin{equation}
    \delta: V \rightarrow W.
\end{equation}
If we write $\delta^*$ for the adjoint to $\delta$ then:
\begin{equation}
    \delta^*: W^* \times V \rightarrow \mathbb{R},
\end{equation}
which is to say:
\begin{equation}
    \delta^*: W^* \rightarrow V^*.
\end{equation}
Which is simply another way of saying that cofunction interpolation is the dual of coefficient interpolation.

\section{External operators}

The delta form is simply one example of an form whose first argument is in a dual space. In fact any operator which outputs a \mintinline{python}|Function| and which is linear in any other arguments it has is a form of similar type. Note that such an operator could have known inputs (e.g. coefficients) with respect to which it is nonlinear, it is only the unknown arguments with respect to which it has to be linear. In this context it is important to distinguish between the inputs to such an operator, which we will call its \emph{operands} and arguments. Arguments may appear in operands, but an operand also contain other UFL expressions. For example an operator $N$ might take some UFL expression as an input, evaluate a neural net and return the output as a \mintinline{python}|Function|. That is $N(e, m)\in V$ for a UFL expression $e$, $m\in\mathbb R^k$ and $V$ some function space. Of course saying that $N(e,m)\rightarrow V$ is the same as saying $N(e, m) \times V^* \rightarrow \mathbb{R}$, so an external operator is also a form.

\section{Form composition}

Recall that a $k$-form of type $V_{k-1}\times\ldots \times V_0 \rightarrow \mathbb{R}$ can be interpreted as an operator $V_{k-1}\times\ldots \times V_1 \rightarrow V_0^*$. In other words, it can mathematically be used anywhere were an object in $V_0^*$ is expected. Similarly, the reflexivity of the vector spaces we work with means that a $k$-form of type $V_{k-1}\times\ldots \times V_0^* \rightarrow \mathbb{R}$ can be interpreted as an operator $V_{k-1}\times\ldots \times V_1 \rightarrow V_0$. That is, it can be used where a value in $V_0$ is expected.

Hitherto, UFL forms have only taken inputs from primal spaces and produced outputs in dual spaces, so composing forms by using the output of one as an input to another was not an available option. However, now that there are forms which produce primal space outputs, and forms which accept dual space inputs, we need to examine the consequences of this.

Consider this simple case:
\begin{minted}{python}
    v_ = ufl.Argument(V.dual, 0)
    f = firedrake.Function(W)
    v = ufl.Argument(V, 0)
    F_1 = ufl.Delta(f, v_) * v * dx
    l = assemble(F_1)
\end{minted}
This is a slightly contrived case, though it could really occur if, for example, $W$ is defined on a different mesh from $V$. The assembly algorithm for \mintinline{python}|l| would be:
\begin{minted}{python}
    tmp = firedrake.Function(V)
    l = assemble(ufl.replace(F_1, {ufl.Delta(f, w_): tmp}))
\end{minted}
I.e., we first evaluate the delta, and then substitute the result into the surrounding form. The type of the temporary variable can be inferred from the arguments to the delta. Consider now the 2-form case:
\begin{minted}{python}
    v_ = ufl.Argument(V.dual, 0)
    w = ufl.Argument(W, 1)
    v = ufl.Argument(V, 0)
    F_2 = ufl.Delta(w, v_) * v * dx
    A = assemble(F_2)
\end{minted}
This results in the following assembly algorithm:
\begin{minted}{python3}
    tmp_1 = assemble(Delta(w, v_)) # tmp_1 is a |V| x |W| matrix.
    tmp_2 = assemble(ufl.replace(F_2, {ufl.Delta(w, v_): ufl.Argument(V, 1)}))
    A = tmp_2 @ tmp_1
\end{minted}
Again the, the types of the temporary variables can be inferred from the arguments of the delta. The delta has 2 arguments and they are in the canonical order, so it is replaced by an argument numbered 1 but in the function space dual to the argument 0 function space. The sequence of the matrix multiplication is similarly inferable from the arguments and the sequence of composition (the delta occurs inside the other form, and not vice versa). The general rule is that the form is replaced by an argument in the dual space to its zero argument, and with an argument number which is the next available argument number in the form into which it is being substituted.

\subsection{Arguments of composite forms}

What are the arguments of \mintinline{python}|F_2|? One might be tempted to say \mintinline{python}|v_|, \mintinline{python}|w|, and \mintinline{python}|v|, but this would characterise \mintinline{python}|F_2| as a 3-form rather than a 2-form. It would also have two different arguments numbered 0 (\mintinline{python}|v_|, and \mintinline{python}|v|), so it would not even be a well-formed form. The answer to this puzzle is that the 0 argument of a form nested into another form is consumed by the substitution. In this case, this means that the arguments of the composite form are \mintinline{python}|v| and \mintinline{python}|w|.

\subsection{No substitution for argument 0}
Consider instead the case where the interpolation happens for the test function. By symmetry with \mintinline{python}|F_2|, one might expect to be able to write:
\begin{minted}{python}
    w_ = ufl.Argument(W.dual, 0)
    v = ufl.Argument(V, 0)
    u = ufl.Argument(V, 1)
    F_3 = u * ufl.Delta(v, w_) * dx
\end{minted}
However this is ill-formed since the delta contains two arguments numbered 0. This difficulty is the result of the fact that form composition only makes sense if we think of forms as operators into the dual space of argument 0. In this context, argument 0 is not a form input at all, but is instead a placeholder for the result of evaluating the form for known values of all of its higher-numbered arguments. However, composition of linear operators works in both directions so we can instead write:
\begin{minted}{python}
    w = ufl.Argument(W, 0)
    v = ufl.Argument(W, 0)
    u = ufl.Argument(V, 1)
    F_4 = ufl.Delta(w, u * v * dx)
\end{minted}
Which is well-formed UFL, and mathematically equivalent to what \mintinline{python}|F_3| attempted to achieve. As one might expect, $F_4: V \rightarrow W^*$.

\section{Action}

The action of a form on a coefficient or argument simply replaces the highest numbered argument with that coefficient. This carries over directly to the dual case. If the highest numbered argument of a form lies in a dual space, then the action is only well defined on a cofunction or coargument from the appropriate space.

Form composition means that it should also be possible to take the action of a form on another form, so long as the space of the highest numbered argument of the first form is dual to the space of the 0 argument to the second form. That is, an alternative definition of \mintinline{python}|F_4| is given by:
\begin{minted}{python}
    w = ufl.Argument(W, 0)
    omega = ufl.Argument(W.dual, 1)
    v = ufl.Argument(V, 0)
    u = ufl.Argument(V, 1)
    F_4 = ufl.action(ufl.Delta(w, omega), u * v * dx)
\end{minted}
Of course this lines up with the assembly strategy for \mintinline{python}|F_4|, which would be:
\begin{minted}{python3}
    tmp_1 = assemble(Delta(w_, ufl.Argument(V.dual, 1))
    tmp_2 = assemble(u * v * dx)
    A = tmp_1 @ tmp_2
\end{minted}
\printbibliography

\end{document}